\documentclass{article}
\usepackage{arxiv}
\usepackage[utf8]{inputenc} 
\usepackage[T1]{fontenc}    
\usepackage{hyperref}       
\usepackage{url}           
\usepackage{booktabs}       
\usepackage{amsfonts}      
\usepackage{nicefrac}      
\usepackage{microtype}      
\usepackage{lipsum}	
\usepackage{graphicx}
\usepackage[numbers, sort&compress]{natbib}
\usepackage{tikz}
\usetikzlibrary{arrows,calc,positioning}
\usepackage{graphicx}
\usepackage{longtable}
\DeclareUnicodeCharacter{FB01}{fi}
\usepackage{caption}
\usepackage{csquotes}
\usepackage{colortbl}
\usepackage{multirow}
\usepackage{tabularx}
\usepackage{multicol}
\usepackage{rotating} 
\usetikzlibrary{arrows,shadows} 
\usepackage[underline=true,rounded corners=false]{pgf-umlsd}
\usepackage{enumitem}
\newlist{steps}{enumerate}{1}
\setlist[steps, 1]{label = Step \arabic*:}
\title{Cryptography approach for Secure Outsourced Data Storage in Cloud Environment}
\author{ \href{https://orcid.org/0000-0001-9305-1269}{\includegraphics[scale=0.06]{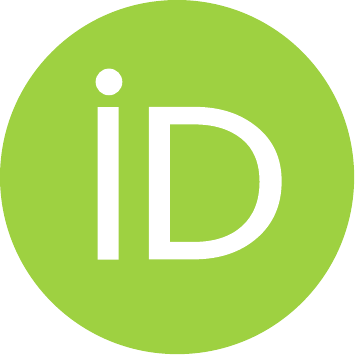}\hspace{1mm}Rishabh~Gupta}\thanks{
} \\
	Department of Computer Science and Engineering\\
	The University of Aizu, Aizuwakamatsu, Japan \\
	\texttt{rishabhgpt66@gmail.com} \\
	\And
\href{https://orcid.org/0000-0002-9689-6387}{\includegraphics[scale=0.06]{orcid.pdf}\hspace{1mm}Deepika~Saxena} \\
	Department of Computer Science and Engineering\\
	The University of Aizu, Aizuwakamatsu, Japan \\
	\texttt{13deepikasaxena@gmail.com} \\
     \And
	\href{https://orcid.org/0000-0002-8053-5050}{\includegraphics[scale=0.06]{orcid.pdf}\hspace{1mm}Ashutosh Kumar~Singh} \\
	Department of Computer Applications\\
	National Institute of Technology\\
	Kurukshetra, India \\
	\texttt{ashutosh@nitkkr.ac.in} \\
}
\renewcommand{\shorttitle}{Secure Outsourced Data Storage in Cloud Environment}
\hypersetup{
pdftitle={A template for the arxiv style},
pdfsubject={q-bio.NC, q-bio.QM},
pdfauthor={Rishabh~Gupta, Ashutosh Kumar~Singh},
pdfkeywords={First keyword, Second keyword, More},
}

\begin{document}
\maketitle

\begin{abstract}
A large amount of data and applications are migrated by researchers, stakeholders, academia, and business organizations to the cloud environment due to its large variety of services, which involve the least maintenance cost, maximum flexibility, and on-demand service for storage, computation, and data distribution intentions. Despite the various characteristics the cloud environment supports, it also faces many challenges. However, data users may not completely trust a cloud environment that is engaged by a third party. Every cloud user always has a prime concern, i.e., security. Numerous methods have been designed to solve the issue of data security during data storage, calculation, and sharing across stakeholders and users. Nevertheless, there is a lack of existing methods that tackle the issue of the security of data when it is stored in a cloud environment. This article presents a precise security method that has handled the security of data while it is being shared and stored in the cloud. These methods have been utilized to lessen security assaults and prevent unauthorized parties from accessing the actual data. The article is concluded with some limitations and recommendations for the future in terms of secure data retention and distribution.
\end{abstract}
\keywords{Cloud Computing \and Data Security \and Cloud Storage \and Privacy Preservation \and Cryptography}
\section{Introduction}
 With its fast data processing and storage capacity, cloud computing is essential for all applications that need high processing costs, such as machine learning for classification \cite{wang2010privacy, ali2015sedasc, sundareswaran2012ensuring, thilakanathan2014secure, dong2014achieving, chaudhary2022privacy, gupta2022privacyArxiv, shen2021privacy, gupta2022iot, gupta2022differentialCICT, gupta2022differentialIEEENetworking, gupta2022differentialAIHC, yang2018differentially, gupta2022differentialNGC, singh2022privacyMultimedia, gupta2022quantum, gupta2022privacyICPC2T, gupta2022differentialJWE}. Cloud computing offers open, on-demand, scalable, and easy-to-use computing services \cite{kumar2023power, singh2022reliable, swain2023ai, singh2023bio, saxena2023performance, saxena2023sustainable, saxena2022high, singh2022cryptography, saxena2022fault, saxena2022ofp, saxena2022intelligent, saxena2022vm, saxena2021secure, saxena2021osc, singh2021quantum, saxena2021op, saxena2021proactive, saxena2020security, kumar2021performance, kumar2020decomposition, kumar2020biphase, kumar2020adaptive, kumar2022security, singh2023privacy}. The cloud server requires a huge quantity of data for computation and sharing among users \cite{singh2020online, gupta2020seli, gupta2019dynamic, singh2019sql, singh2018data, singh2019secure, kumar2018long, chhabra2018probabilistic, makkar2022secureiiot, singh2023metaheuristic, yadav2022survey, singh2023secure}. These data are obtained from different data owners. Every data owner has its data containing sensitive data such as personal images, social media data, and medical records \cite{Utkarsh2022Data, tripathi2020review, chauhan2020survey, tripathi2021hedcm, deepika2020review, Atmapoojya2022Data, Pardhan2021CPABE, Rani2021DL}. 
 Such sensitive data is transmitted to the cloud service provider for keeping and processing purposes \cite{gupta2022privacyIJCAET, saxena2021survey, gupta2021dataArxiv, chhabra2022comprehensive, chhabra2016dynamic, swain2022efficient}. The data owners lose their rights, and also do not know about who is accessing outsourced data \cite{gupta2020mlpam, singh2022privacy}. But, after outsourced data, different organizations or any adversary can access the data since the cloud service provider holds all data on its servers and it is the third party \cite{chhabra2023secure, chhabra2020security}. The data owner may not trust the cloud server \cite{zaghloul2019p, sun2021share}. Cloud computing suffers from various security problems, which are widely discussed topics \cite{ali2015security}. In fact, cloud computing offers many tools but also has critical security concerns. Therefore, to preserve data privacy from others, owners first convert their data before passing data to another entity. Although the data is converted using existing approaches, such as differential privacy \cite{wei2020federated, gong2020privacy, chamikara2020privacy, fan2020privacy, wu2022adaptive, wang2018dp3, ma2021data, du2018differential, zhang2017efficient, bao2021privacy, kim2019udipp, wu2020privacy}. To protect the actual information of the owners, the existing approach produces the noise using a probability-based distribution function. The generated noise is added to the actual information and acquires the noise-added data which is stored on the cloud server and shared across all the entities. But this approach does not trade-off between accuracy and privacy. To resolve the above challenge, a cryptography approach for secure outsourced data storage in the cloud environment is proposed for secure computation and distribution among entities. The proposed scheme adopts a cryptography approach to encrypt the owners' data. Data is secured by encryption and decryption processes. So that the adversary is unable to understand the actual data of each owner. In this proposed model, it is considered that the data owners are not communicating with each other. The cloud server is honest-but-curious (non-colluding), following the protocol strictly. The data in encrypted form is forwarded to the cloud server that stores it. Cloud server performs computation on the received data and shares it among users whenever they send the request for the owners’ data. Users decrypt the data that are obtained from the cloud service provider and acquire the actual information.
Therefore, the proposed model stores and shares the data in such a way that doesn't reveal the data owners' data. 
\section{Related Work}
To assess the suitability of fog computing, a three-tier architecture is presented in \cite{sarkar2015assessment}, which transmits the data from terminal nodes to the cloud through the fog nodes.  On the basis of this model, the authors described the mathematical disparity between the traditional cloud computing paradigm and the fog computing paradigm for different renewable and non-renewable energy resources and related costs. A case study was conducted and the result revealed that fog computing outperforms cloud computing in the sense of the Internet of Things ( IoT), with a large number of latency-sensitive applications. The proposed architecture reduced latency, energy consumption, and carbon dioxide emissions but suffered from limited data sharing.

In the current storage schema, the data of the users are stored entirely on cloud servers. By doing so, users lose their right to data protection and face the possibility of privacy leakage. 
To preserve the privacy of data, a fog computing-driven three-layer storage framework was devised by Wang et al. \cite{wang2018three}. The suggested system has the potential to make the most of cloud storage while still preserving data privacy. A distribution method based on risk mitigation is used to spread the components among various cloud storage platforms. In addition, the Hash-Solomon Code algorithm was also designed to divide data into different sections. Each component can be put on the cloud, fog, and local machines and reconstructed from plaintext. The privacy protection of sensitive data is tackled by fog computing-based data collection methods with exceptional performance. However, user revocation was not considered.

The Server-aided Network Topology (SNT) system and the Fully-connected Network Topology (FNT) system, both of which are based on connections to SNT and FNT servers, were introduced by Phong and Phuong \cite{phuong2019privacy} for stochastic gradient descent (SGD) protection. The SGD or its derivatives can be used by several machine learning trainers in these systems over the combined dataset without sharing the local dataset of each trainer. The clients communicate the locally derived weights for the model to the server, which then aggregates the incoming weights and transmits them back to the clients. Applying privacy-preserving weight transfer to network training in the encrypted domain can offer robustness against model extraction attempts. Such an approach's benefit is that it withstands updated loss and doesn't need frequent synchronization. The experiments were conducted using various datasets, and results showed that their system outperformed the existing system in terms of learning accuracy. These systems are also effective in terms of calculation \& communication. The developed systems, which utilized weight parameters rather than gradient parameters, had accuracy close to SGD.


A privacy-preserving price e-negotiation (3PEN) protocol is introduced in \cite{liu2019quantum} that protects the prices of the seller and buyer. A secure multi-party computation (SMC) technique was employed for the input data encryption. The oracle and qubit comparator was applied to acquire the final state, depicting the two-party’s price results. It presents the results of the prices of the two parties' comparisons, and counting is done to total the number of products that satisfy the trading requirements (i.e., the number of times the buyer's offer exceeds the seller's asking price by a predetermined amount). This prevents any confidential information about Alice's prices from being discovered. The suggested protocol offers higher security and significantly lowers the likelihood of an external eavesdropper attack than the traditional channel while increasing computing costs.
\begin{table}[htbp]
	\caption{A capsulization of cryptography-based models}	
  \label{TableExp1}
		\small\addtolength{\tabcolsep}{-4pt}
  \begin{tabular} {|p{1.8cm}|p{4.3cm}|p{3.3cm}|p{3.3cm}|p{2.3cm}|}
\hline
Model/Scheme /Framework & Workflow & Implementation & Outcomes & Drawbacks \&
Future Scope \\
\hline  \hline
 A fog computing-based network model for data service \cite{sarkar2015assessment} & \vspace{-0.3cm} 
        \begin{itemize}[leftmargin=0.3cm,label=\textbullet]
            \item The network model of fog computing was constructed to exchange traffic pattern
            \item A case study was performed with traffic generated from the 100 highest-populated cities 
        \end{itemize} & 
         \vspace{-0.3cm} 
        \begin{itemize}[leftmargin=0.3cm,label=\textbullet]
        \item The experimental results performed range from 2 to 10 terminal nodes
        \item The ratio of the total bytes transmitted to the fog computing is [0.05, 0.75]
        \end{itemize}
         &
      \vspace{-0.3cm} 
        \begin{itemize}[leftmargin=0.3cm,label=\textbullet]
            \item Reduced 50\% service latency due to fog computing
            \item Less computational and storage overhead 
        \end{itemize} 
                 &
      \vspace{-0.3cm} 
        \begin{itemize}[leftmargin=0.3cm,label=\textbullet]
            \item This work can be extended by considering the heterogeneous devices
            \end{itemize} \\
        \hline \hline                
A cloud storage scheme for data privacy \cite{wang2018three} & \vspace{-0.3cm} 
        \begin{itemize}[leftmargin=0.3cm,label=\textbullet]
            \item The data was encoded using hash transformation
            \item  The distribution proportion was computed to divide the data into small parts
        \end{itemize} & 
         \vspace{-0.3cm} 
        \begin{itemize}[leftmargin=0.3cm,label=\textbullet]
        \item The experiments were performed based on ‘one more block’ principle
        \item Audio (.MP3, 84.2 MB), Picture (.NEF, 24.3 MB), and Video (.RMVB, 615 MB) were used for experiments
        \end{itemize}
         &
      \vspace{-0.3cm} 
        \begin{itemize}[leftmargin=0.3cm,label=\textbullet]
            \item It reduced the lower servers' storage pressure
            \item Safety analysis proves that this framework is secure, feasible, and efficient
        \end{itemize} 
                 &
      \vspace{-0.3cm} 
        \begin{itemize}[leftmargin=0.3cm,label=\textbullet]
            \item A fraction of data gets exposed to each outsourcing cloud server
            \end{itemize} \\
        \hline \hline
A secure sharing network weights system driven on multilayered neural network \cite{phuong2019privacy} & \vspace{-0.3cm} 
        \begin{itemize}[leftmargin=0.3cm,label=\textbullet]
            \item The security of input data was performed via symmetric encryption
            \item SGD computed the gradient parameters
            \end{itemize} & 
      \vspace{-0.3cm} 
        \begin{itemize}[leftmargin=0.3cm,label=\textbullet]
            \item Breast Cancer, Skin/non-Skin, and MNIST datasets were adopted to evaluate this framework 
            \item The experimental results were achieved in the python programming language
        \end{itemize} & 
      \vspace{-0.3cm} 
        \begin{itemize}[leftmargin=0.3cm,label=\textbullet]
            \item Maintain the accuracy of deep learning at its highest level 
            \item According to security analysis, the proposed framework is more computationally efficient  
              \end{itemize} & 
      \vspace{-0.3cm} 
        \begin{itemize}[leftmargin=0.3cm,label=\textbullet]
            \item The privacy of the output is not being considered 
     \item Low efficiency and lack of privacy concerns are caused by updated weight 
                \end{itemize} \\
        \hline \hline
 A secure multi-party computation-based e-negotiation protocol for price protection \cite{liu2019quantum} & \vspace{-0.3cm} 
        \begin{itemize}[leftmargin=0.3cm,label=\textbullet]
            \item The two-party’s prices were acquired by using a qubit comparator
            \item The corresponding state was prepared with the assistance of Oracle operations
            \end{itemize} & 
      \vspace{-0.3cm} 
        \begin{itemize}[leftmargin=0.3cm,label=\textbullet]
            \item Not Available
        \end{itemize} & 
      \vspace{-0.3cm} 
        \begin{itemize}[leftmargin=0.3cm,label=\textbullet]
            \item The proposed model is more secure according to privacy analysis 
            \item Less communication cost
              \end{itemize} & 
      \vspace{-0.3cm} 
        \begin{itemize}[leftmargin=0.3cm,label=\textbullet]
            \item Only one-to-one communication (between seller and buyer)
            \item This work can be extended by using multi-party price e-negotiation
                \end{itemize} \\
        \hline \hline    
 OU cryptosystem-based privacy-preserving protocol for data classification \cite{chai2020improvement} & \vspace{-0.3cm} 
        \begin{itemize}[leftmargin=0.3cm,label=\textbullet]
            \item The data was encrypted using OU encryption algorithm
            \item Users acquired the secure classification results by applying an outsourced classification protocol 
            \end{itemize} & 
      \vspace{-0.3cm} 
        \begin{itemize}[leftmargin=0.3cm,label=\textbullet]
            \item The MIRACL library was used for simulations
            \item The Naive Bayes classifier was trained using titanic, thyroid, breast cancer, and heart data sets
        \end{itemize} & 
      \vspace{-0.3cm} 
        \begin{itemize}[leftmargin=0.3cm,label=\textbullet]
            \item Increased efficiency due to OU cryptosystem  
            \item The computation cost is modulus $n$, which is less than the cost of the Paillier cryptosystem
              \end{itemize} & 
      \vspace{-0.3cm} 
        \begin{itemize}[leftmargin=0.3cm,label=\textbullet]
            \item A large number of calculations makes the system complex
            \item The system's efficiency can be enhanced by applying the optimization mechanisms
                \end{itemize} \\
        \hline        
\end{tabular}
\end{table}

Due to resource limitations, classifier owners employ outsourced classification services to move their classifiers to remote servers, where users request services of classification. To protect the classifier, Chai et al. \cite{chai2020improvement} devised an outsourced classification protection scheme having less computational and communication overhead. The authors utilized a substitutive OU cryptosystem to mitigate the bandwidth consumption. The proposed scheme resists the substitution-then-comparison (STC) attack, ensuring that users can accurately receive the classification outcome without disclosing the classifier's private, with limited data sharing.
Table I presents a thumbnail of relevant models based on the classical mechanism containing potential details.
\section{System Model}
The system model consists of three entities, such as Data Owners ($DO$), Users ($U$), and Cloud Service Provider ($CSP$).
	\begin{figure*}[!htbp]
		\centering{\includegraphics[scale=.35]{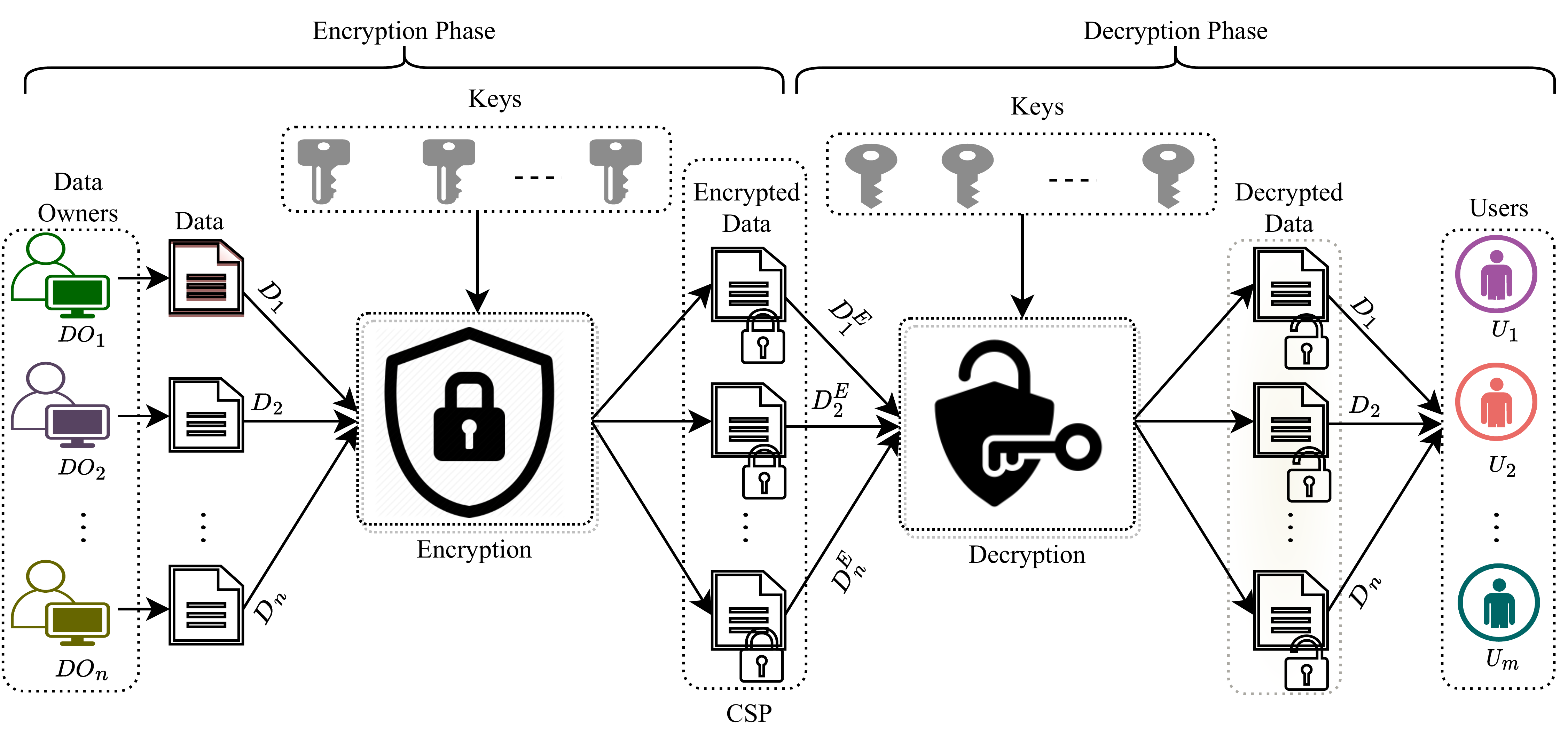}}
		\caption{Schematic representation of cryptography-driven model}
		\label{fig:Systemmodel}
\end{figure*}
\par
Fig. 1 presents the building block of the cryptography technique, which consists of symmetric and asymmetric approaches. In symmetric approach, each data owner $DO_{1}$, $DO_{2}$, $\dots$, $DO_{n}$ and user $U_{1}$, $U_{2}$, $\dots$, $U_{m}$ has secret keys $S_{\mathcal{K}1}$, $S_{\mathcal{K}2}$, $\dots$, $S_{\mathcal{K}n}$. $DO_{1}$, $DO_{2}$, $\dots$, $DO_{n}$ encrypts data $D_{1}$, $D_{2}$, $\dots$, $D_{n}$ with their secret keys $S_{\mathcal{K}1}$, $S_{\mathcal{K}2}$, $\dots$, $S_{\mathcal{K}n}$, respectively, and obtained encrypted data $D_{1}^{E}$, $D_{2}^{E}$, $\dots$, $D_{n}^{E}$. The produced data are passed to the users $U_{1}$, $U_{2}$, $\dots$, $U_{m}$ through $CSP$ for utilization purposes. $U_{1}$, $U_{2}$, $\dots$, $U_{m}$ decrypt the obtained encrypted data $D_{1}^{E}$, $D_{2}^{E}$, $\dots$, $D_{n}^{E}$ with their secret keys $S_{\mathcal{K}1}$, $S_{\mathcal{K}2}$, $\dots$, $S_{\mathcal{K}n}$ and acquire the plain documents $D_{1}$, $D_{2}$, $\dots$, $D_{n}$. Similarly, in asymmetric approach, each data owner $DO_{1}$, $DO_{2}$, $\dots$, $DO_{n}$ has data $D_{1}$, $D_{2}$, $\dots$, $D_{n}$, and public keys $P_{\mathcal{B}1}$, $P_{\mathcal{B}2}$, $\dots$, $P_{\mathcal{B}n}$. $DO_{1}$, $DO_{2}$, $\dots$, $DO_{n}$ encrypts data $D_{1}$, $D_{2}$, $\dots$, $D_{n}$ with their $P_{\mathcal{B}1}$, $P_{\mathcal{B}2}$, $\dots$, $P_{\mathcal{B}n}$, respectively, and acquired encrypted data $D_{1}^{E}$, $D_{2}^{E}$, $\dots$, $D_{n}^{E}$. The acquired data are transferred to the users $U_{1}$, $U_{2}$, $\dots$, $U_{m}$ through $CSP$ for various purposes. $U_{1}$, $U_{2}$, $\dots$, $U_{m}$ decrypt the acquired encrypted data $D_{1}^{E}$, $D_{2}^{E}$, $\dots$, $D_{n}^{E}$ with their private keys $P_{\mathcal{V}1}$, $P_{\mathcal{V}2}$, $\dots$, $P_{\mathcal{V}m}$ and obtain the plain informations $D_{1}$, $D_{2}$, $\dots$, $D_{n}$.
\section{Data Encryption and Decryption}
Let data $D_{1}$, $D_{2}$, $\dots$, $D_{n}$ $\in$ $\mathbb{D}$ that are protected by encryption and decryption process of symmetric and asymmetric cryptography techniques using sets of public keys ($P_{\mathcal{B}}$), secret keys ($S_{\mathcal{K}}$), and private keys ($P_{\mathcal{V}}$). The symmetric cryptography technique maps $\Omega_{E}$ : $D_{i}$ $\times$ $S_{\mathcal{K}}$ $\rightarrow$ $D_{i}^{E}$, and $\Lambda_{D}$; $D_{i}^{E}$ $\times$ $S_{\mathcal{K}}$ $\rightarrow$ $D_{i}$, such that $\Lambda_{D}$$\Omega_{E}$($\mathcal{D}_{i}$, $S_{k}$) = $\mathcal{D}_{i}$; whereas $\Omega_{E}$ and $\Lambda_{D}$; are the encryption and decryption operations. While the asymmetric cryptography technique maps $\Omega_{E}$ : $D_{i}$ $\times$ $P_{\mathcal{B}}$ $\rightarrow$ $D_{i}^{E}$, and $\Lambda_{D}$; $D_{i}^{E}$ $\times$ $P_{\mathcal{V}}$ $\rightarrow$ $D_{i}$, such that $\Lambda_{D}$$\Omega_{E}$($\mathcal{D}_{i}$, $\mathcal{P}_{b}$) = $\mathcal{D}_{i}$; $\forall$ $\mathcal{D}_{i}$ $\in$ $D_{i}$, $\mathcal{S}_{k}$ $\in$ $S_{\mathcal{K}}$, $\mathcal{P}_{b}$ $\in$ $P_{\mathcal{B}}$, $\mathcal{P}_{v}$ $\in$ $P_{\mathcal{V}}$ where $\mathcal{D}_{i}^{E}$ $\in$ $D_{i}^{E}$. 
\par
The symmetric cryptography technique ($\Omega_{E}$, $\Lambda_{D}$, $D_{i}$, $D_{i}^{E}$, $S_{\mathcal{K}}$) comprises three operations: Key Generation, Encryption, and Decryption, which are defined as:
\begin{enumerate}
\item 
\textbf{Key Generation} ($K_{G}^{SY}$ ($\mathcal{CG}$)): This function generates the secret key ($\mathcal{S}_{k}$) using Eq. (1), which is utilized to encrypt and decrypt the data. 
\begin{equation}
 \mathcal{S}_{k} = K_{G}^{SY}(\mathcal{CG}) \forall \mathcal{S}_{k} \in S_{\mathcal{K}}
\end{equation}
\item \textbf{Encryption} ($\Omega_{E}$): This function ($\Omega_{E}$ : $\mathcal{D}_{i}$ $\times$ $\mathcal{S}_{k}$ $\rightarrow$ $\mathcal{D}_{i}^{E}$) performs the encryption task on the actual data $\mathcal{D}_{i}$ using the secret key ($\mathcal{S}_{k}$) and procures the encrypted data $\mathcal{D}_{i}^{E}$, as given in Eq. (2).
\begin{equation}
 \mathcal{D}_{i}^{E} = \Omega_{E}(\mathcal{D}_{i}, \mathcal{S}_{k}) \forall \quad \mathcal{D}_{i} \in D_{i} \wedge \mathcal{S}_{k} \in S_{\mathcal{K}} \wedge \mathcal{D}_{i}^{E} \in D_{i}^{E}  
\end{equation}
\item \textbf{Decryption} ($\Lambda_{D}$): This function ($\Lambda_{D}$; $\mathcal{D}_{i}^{E}$ $\times$ $\mathcal{S}_{k}$ $\rightarrow$ $\mathcal{D}_{i}$) takes the encrypted data $\mathcal{D}_{i}^{E}$ and $\mathcal{S}_{k}$ as input and provides the actual data $\mathcal{D}_{i}$ as an output using Eq. (3).
\begin{equation}
\mathcal{D}_{i} = \Lambda_{D}(\mathcal{D}_{i}^{E}, \mathcal{S}_{k}) \forall \mathcal{D}_{i}^{E} \in D_{i}^{E} \wedge \mathcal{S}_{k} \in S_{\mathcal{K}} \wedge \mathcal{D}_{i} \in D_{i}    
\end{equation}
    \end{enumerate}
The asymmetric cryptography technique ($\Omega_{E}$, $\Lambda_{D}$, $D_{i}$, $D_{i}^{E}$, $P_{\mathcal{B}}$, $P_{\mathcal{V}}$) contains three operations that are described as:
\begin{enumerate}
\item \textbf{Key Generation} ($K_{G}^{AS}$ ($\mathcal{CG}$)): This operation generates the keys $\mathcal{P}_{b}$, and $\mathcal{P}_{v}$ for encryption and decryption, respectively, using Eq. (4).
\begin{equation}
    \mathcal{P}_{b}, \mathcal{P}_{V} = K_{G}^{AS} (\mathcal{CG}) \forall \mathcal{P}_{b} \in P_\mathcal{B} \wedge \mathcal{P}_{v} \in P_\mathcal{V}   
\end{equation}
\item \textbf{Encryption} ($\Omega_{E}$): This function ($\Omega_{E}$ : $\mathcal{D}_{i}$ $\times$ $\mathcal{P}_{b}$ $\rightarrow$ $\mathcal{D}_{i}^{E}$) encrypt the actual information $\mathcal{D}_{i}$ with $\mathcal{P}_{b}$ keys and provides the encrypted data $\mathcal{D}_{i}^{E}$ by applying Eq. (5).
\begin{equation}
\mathcal{D}_{i}^{E} = \Omega_{E}(\mathcal{D}_{i}, \mathcal{P}_{b}) \forall \quad \mathcal{D}_{i} \in D_{i} \wedge \mathcal{P}_{b} \in P_{\mathcal{B}} \wedge \mathcal{D}_{i}^{E} \in D_{i}^{E}  
\end{equation}
\item \textbf{Decryption} ($\Lambda_{D}$): This function ($\Lambda_{D}$; $\mathcal{D}_{i}^{E}$ $\times$ $\mathcal{P}_{v}$ $\rightarrow$ $\mathcal{D}_{i}$) takes the encrypted data $\mathcal{D}_{i}^{E}$ and $\mathcal{P}_{v}$ as input and provides the actual data $\mathcal{D}_{i}$ as an output using Eq. (6).
\begin{equation}
\mathcal{D}_{i} = \Lambda_{D}(\mathcal{D}_{i}^{E}, \mathcal{P}_{v}) \forall \mathcal{D}_{i}^{E} \in D_{i}^{E} \wedge \mathcal{P}_{v} \in P_{\mathcal{V}} \wedge \mathcal{D}_{i} \in D_{i}    
\end{equation}
    \end{enumerate}
\section{Conclusion}
Data security is an especially challenging issue in the cloud computing environment for storage and information sharing. This work focuses on the security of data that has been outsourced by various data owners. It provides a succinct and concise overview of the methods used for outsourcing data security. The system model, data encryption, and decryption processes are discussed, followed by a comparison of various existing works.  A comprehensive data security method that can secure the data and effectively use it for machine learning is required in light of the aforementioned evaluation and new obstacles. This work can be extended by designing a more efficient privacy-preserving mechanism to secure the data for various owners.
\bibliographystyle{unsrt}
\bibliography{reference.bbl} 
\end{document}